\documentclass[a4paper,cits]{article}
\usepackage{amsmath,amssymb,amsfonts,amsthm,latexsym} 

\usepackage{pos}
\usepackage[normalem]{ulem}
\usepackage[utf8]{inputenc}
\usepackage[T1]{fontenc}
\usepackage{cancel}
\usepackage{graphicx}
\usepackage{latexsym}
\usepackage{bbm}
\usepackage{epsfig}
\usepackage{epstopdf}
\usepackage{subcaption}
\usepackage{color}
\usepackage{comment}
\usepackage{slashed}
\usepackage{placeins}
\usepackage{cleveref}
\usepackage{setspace}
\usepackage{url}
\usepackage{tabularx}
\usepackage{xcolor}
\usepackage{tikz,wrapfig}
\usepackage{enumitem}
\usetikzlibrary{fadings,calc,decorations.markings}

\makeatletter
\newsavebox{\@brx}
\newcommand{\llangle}[1][]{\savebox{\@brx}{\(\m@th{#1\langle}\)}%
  \mathopen{\copy\@brx\kern-0.5\wd\@brx\usebox{\@brx}}}
\newcommand{\rrangle}[1][]{\savebox{\@brx}{\(\m@th{#1\rangle}\)}%
  \mathclose{\copy\@brx\kern-0.5\wd\@brx\usebox{\@brx}}}
\makeatother

\makeatletter
\def\namedlabel#1#2{\begingroup
    #2%
    \def\@currentlabel{#2}%
    \phantomsection\label{#1}\endgroup
}
\makeatother

\newcommand{\ZE}{Z_\textrm{E}}

\newcommand{\als}{\alpha_{\text{s}}}

\title{The static force from generalized Wilson loops}
\ShortTitle{The static force from generalized Wilson loops}

\author*[a,1]{Viljami Leino}
\author[a,b,c,1]{Nora Brambilla}
\author[d]{Owe Philipsen}
\author[d,e]{Christian Reisinger}
\author[a,1]{Antonio Vairo}
\author[d,e]{Marc Wagner}

\affiliation[a]{Physik Department, Technische Universit\"at M\"unchen,\\
James-Franck-Strasse 1, 85748 Garching, Germany}
\affiliation[b]{Institute for Advanced Study, Technische Universit\"at M\"unchen,\\
Lichtenbergstrasse 2 a, 85748 Garching, Germany}
\affiliation[c]{Munich Data Science Institute, Technische Universit\"at M\"unchen, \\
Walther-von-Dyck-Strasse 10, 85748 Garching, Germany}
\affiliation[d]{Goethe Universit\"at Frankfurt am Main, Institut für Theoretische Physik, \\ Max-von-Laue-Str.\ 1, 60438 Frankfurt am Main, Germany}
\affiliation[e]{Helmholtz Research Academy Hesse for FAIR, Campus Riedberg, \\ Max-von-Laue-Stra{\ss}e 12, D-60438 Frankfurt am Main, Germany}

\emailAdd{viljami.leino@tum.de}
\emailAdd{nora.brambilla@ph.tum.de}
\emailAdd{philipsen@itp.uni-frankfurt.de}
\emailAdd{reisinger@itp.uni-frankfurt.de}
\emailAdd{antonio.vairo@ph.tum.de}
\emailAdd{mwagner@itp.uni-frankfurt.de}

\note{For TUMQCD collaboration}

\abstract{Recently a method to compute the static force with lattice gauge theory using an insertion of a chromoelectric field into a Wilson loop was proposed.
          We explore this method using the multilevel algorithm and
          discuss the renormalization of the chromoelectric field on the lattice.
		 }

\FullConference{%
 The 38th International Symposium on Lattice Field Theory, LATTICE2021
  26th-30th July, 2021
  Zoom/Gather@Massachusetts Institute of Technology
}

\begin{document}
\maketitle

\section{Introduction}
The potential between a static quark-antiquark pair $V(r)$ is one of the most commonly studied quantities in QCD. 
At small separations $r$ the static potential can be calculated in a weak coupling expansion. 
The perturbative expression of the static energy is known at N$^3$LL accuracy and can be combined with high precision measurements of the same quantity in lattice QCD. This allows an accurate extraction of 
the strong coupling $\alpha_{\rm s}$, which is competitive with lattice determinations from different observables \cite{Aoki:2019cca}.

In a lattice regularization, the static potential comes with a linear divergence of order $\alpha_{\rm s}(1/a)/a$ (with $a$ denoting the lattice spacing) also referred to as self-energy.
The self-energy vanishes in dimensional regularization. 
However, the perturbative expression for $V(r)$ in dimensional regularization is affected by a renormalon ambiguity of order $\Lambda_\text{QCD}$~\cite{Pineda:1998id,Hoang:1998nz}.
Both the renormalon and the self energy can be absorbed into an additive constant.
This constant disappears, when considering the static force $F(r) = \partial_r V(r)$.
The static force encodes the shape of $V(r)$ and carries all the physical information needed to extract $\als$, while being finite and renormalon free.

A precise computation of the static force on the lattice can be challenging.
While the traditional way of computing the static force by taking finite differences of the static potential \cite{Necco:2001xg,Necco:2001gh} can be efficient in quenched lattice QCD,
the lattice data points for $V(r)$ might be too sparse in full QCD for a reliable extraction of the static force either via finite differences or
via interpolation \cite{Bazavov:2014soa}.
Such problems can be avoided by using a recently suggested method \cite{Vairo:2015vgb,Vairo:2016pxb} based on Ref.\ \cite{Brambilla:2000gk}: The force between a static quark and a static antiquark is computed directly from the expectation value of a Wilson loop with a chromoelectric field inserted in one of the temporal Wilson lines.

In this paper we carry out a quenched lattice QCD computation of the static force using this new method.
We discuss, how to obtain the static force either from Wilson or Polyakov loops with chromoelectric field insertions.
Both approaches yield consistent results. The corresponding systematic errors are, however, different.
We also address the issue that the discretized chromoelectric field has a slow convergence towards the continuum limit, unless a multiplicative renormalization factor is introduced.

This conference contribution is organized in the following way: We introduce our observables in section~\ref{sec2} and the simulation setup in section~\ref{sec:setup}.
The renormalization of this definition of the static force is discussed in section~\ref{sec:reno}.
Numerical results for the static force are presented in section~\ref{sec:res}.
Further details can be found in our recent publication \cite{Brambilla:2021wqs}.
Results obtained at an early stage of this project were discussed at a previous edition of the lattice conference~\cite{Brambilla:2019zqc}.

\section{Observables and their discretizations}\label{sec2}

The force $F(r)$ between a static quark and an antiquark is the derivative of the static potential $V(r)$,
\begin{eqnarray}\label{eq:fdv}
F(r) = \partial_r V(r) .
\end{eqnarray}
$V(r)$ is related to rectangular Wilson loops with large temporal extent $T$ and arbitrary spatial extent $r$,
\begin{eqnarray}
\label{EQN764} e^{-V(r) T} \sim {\rm Tr}\{{\rm P} \, W_{r \times T}\} = {\rm Tr}\bigg\{{\rm P} \, \exp\bigg(i g \oint_{r \times T} dx_\mu \, A_\mu(x)\bigg)\bigg\} .
\end{eqnarray}

On a lattice the Wilson loop $W_{r \times T}$ is discretized by a product of link variables $U_\mu(x) = e^{i a g A_\mu(x)}$. 
The static potential is typically extracted via
\begin{eqnarray}
\label{EQN_V_lat} V(r,a) = \lim_{T \rightarrow \infty}  V_\text{eff}(r,T,a) \quad , \quad V_\text{eff}(r,T,a) = -\frac{1}{a} \ln \frac{\langle \textrm{Tr}\{{\rm P} \, W_{r \times (T+a)}\}\rangle}{\langle \textrm{Tr}\{{\rm P} \, W_{r \times T}\}\rangle} .
\end{eqnarray}
The static force can the be obtained in a straightforward way from the static potential by using a discrete derivative, e.g.\
\begin{eqnarray}
\label{EQN_F_unimproved} F_{\partial V}(r,a) = \frac{V(r+a,a) - V(r-a,a)}{2 a} .
\end{eqnarray}

Alternatively, one can compute the static force using a method proposed in Refs.\ \cite{Vairo:2015vgb,Vairo:2016pxb},
\begin{eqnarray}
\label{EQN_F} F(r) = \lim_{T \rightarrow \infty} -i \frac{\langle {\rm Tr}\{{\rm P} \, W_{r \times T} \,\hat{\mathbf{r}} \cdot g \mathbf{E}(\mathbf{r},t^\ast)\}\rangle}{\langle \textrm{Tr}\{{\rm P} \, W_{r \times T}\} \rangle} \,.
\end{eqnarray}
Here $\hat{\mathbf{r}}$ is the spatial direction of the separation of the static quark-antiquark pair
and $\mathbf{E}(\mathbf{r},t^\ast)$ denotes the chromoelectric field inserted on one of the temporal Wilson lines at a fixed time $t^\ast$. 
The chromoelectric field components are defined as $E_j(x) = F_{j 0}(x)$ in terms of the non-Abelian field strength tensor. 
We note that $\langle {\rm Tr}\{{\rm P} \, W_{r \times T} \, g E_j({\bf r},t^*)\} \rangle$ depends on $t^\ast$. This dependence, however, disappears in the limit $T \rightarrow \infty$, if $t^\ast$ is kept constant.

The lattice formulation of the right hand side of Eq.\ \eqref{EQN_F} requires a discretized field insertion $E_j$. We use two different symmetric formulations, a so-called butterfly
\begin{eqnarray}
\label{EQN_butterfly} \Pi_{j 0} = \frac{P_{j,0}+P_{0,-j}}{2}
\end{eqnarray}  
or a cloverleaf
\begin{eqnarray}
\label{EQN_cloverleaf} \Pi_{j 0} = \frac{P_{j,0} + P_{0,-j} + P_{-j,-0} + P_{-0,j}}{4}
\end{eqnarray}
of plaquettes $P_{\mu,\nu} = 1 + i a^2 g F_{\mu \nu} + \mathcal{O}(a^4) = U_\mu(x) U_\nu(x+\hat{\mu}) U^\dagger_\mu(x+\hat{\nu}) U_\nu^\dagger(x)$~\cite{Bali:1997am}. In both cases the chromoelectric field is given by
\begin{eqnarray}
g E_j = \frac{\Pi_{j 0} - \Pi^\dagger_{j 0}}{2 i a^2} + \mathcal{O}(a^2) .
\end{eqnarray}
Moreover, it is convenient to define an effective force and to extract the static force via
\begin{equation}
\label{eq:f1} F_E(r,a) = \lim_{T \rightarrow \infty} F_{E,\text{eff}}(r,T,a) \quad , \quad F_{E,\text{eff}}(r,T,a) = -i \frac{\langle \textrm{Tr}\{{\rm P} \, W_{r \times T} \, \hat{\mathbf{r}} \cdot g \mathbf{E}(\mathbf{r},t^*)\}\rangle}{\langle \textrm{Tr}\{{\rm P} \, W_{r\times T}\}\rangle} .
\end{equation}
In the following we insert the chromoelectric field $\mathbf{E}(\mathbf{r},t^*)$ exclusively at $t^* = 0$. 
While the continuum formulation~\eqref{EQN_F} is independent of $t^*$, the choice $t^* = 0$ maximizes the the distance to both  temporal boundaries of the Wilson loop and, thus,
should lead to a stronger suppression of excitations and, consequently, to a clearer signal.
Furthermore, we improve the lattice results for the static force using perturbation theory at tree-level.
This is done by redefining the discrete lattice separations $r$ in such a way that the right-hand-side of Eq.~\eqref{EQN_F} in lattice tree-level perturbation theory and in continuum tree-level perturbation theory are identical.

Instead of Wilson loops one can also consider correlation functions of Polyakov loops to compute the static potential as well as the static force.
A Polyakov loop is defined as the normalized trace of a closed temporal Wilson line winding around the periodic temporal direction of extent $T$,
\begin{eqnarray}
L({\bf x}) = \frac{1}{N_c} {\rm Tr} \bigg\{{\rm P} \, \exp\bigg(i g \int_0^T dt \, A_0(x)\bigg)\bigg\} .
\end{eqnarray}
For the static force a Polyakov loop with a chromoelectric field insertion is needed,
\begin{eqnarray}
L_E(\mathbf{r}) = \frac{1}{N_c} {\rm Tr}\bigg\{{\rm P} \, \exp\bigg(i g \int_{t^*}^T dt \, A_0(x)\bigg) \hat{\mathbf{r}} \cdot g \mathbf{E}(\mathbf{r},t^*) {\rm P} \, \exp\bigg(i g \int_0^{t^*} dt \, A_0(x)\bigg)\bigg\} .
\end{eqnarray}
The analogue of Eq.~\eqref{EQN_F} then reads
\begin{eqnarray}
\label{EQN_F_LL} F(r) = \lim_{T \rightarrow \infty} -i \frac{\langle L^\dag({\bf 0}) L_E(\mathbf{r}) \rangle}{\langle L^\dag({\bf 0}) L(\mathbf{r}) \rangle} .
\end{eqnarray}

\section{\label{sec:setup}Simulation setup}

We discretize the SU(3) Yang-Mills theory using the standard Wilson plaquette action.
We carried out simulations with the multilevel algorithm~\cite{Luscher:2001up}, where we performed the updates using the heatbath and overrelaxation algorithms.
We generated three ensembles with lattice spacings
$a=0.060 \, \text{fm}$, $a=0.048 \, \text{fm}$ and $a=0.040 \, \text{fm}$, which we refer to as ensembles $A$, $B$, and $C$, respectively.
The lattice spacing in units of the scale $r_0$ is related to the gauge coupling $\beta$ with a parameterization from Ref.\ \cite{Necco:2001xg}.
The full set of simulation parameters for the ensembles $A$, $B$, and $C$ can be found in Ref.\ \cite{Brambilla:2021wqs}.

To improve the ground state overlaps generated by the spatial Wilson lines in the Wilson loops, 
we use APE smeared spatial links with $\alpha_\textrm{APE} = 0.5$ and $N_\textrm{APE} = 50$ smearing steps for the Wilson loops (for detailed
equations see e.g.\ Ref.\ \cite{Jansen:2008si}).

\section{Renormalization}\label{sec:reno}

On the lattice the two definitions of the static force, $F_E(r,a)$ and $F_{\partial V}(r,a)$ (\eqref{eq:f1} and \eqref{EQN_F_unimproved}, respectively), 
lead to significantly different discretization errors. In other words, the convergence of these observables to the continuum result is quite different.
Such differences are expected, because it is known that observables involving components of the field strength tensor often exhibit sizable discretization errors at values of the gauge coupling typically used in numerical simulations.
The reason is the slow convergence of lattice perturbation theory, when expanded in the bare coupling \cite{Lepage:1992xa}. 
To reduce discretization errors arising from chromoelectric 
and chromomagnetic field insertions, one can use multiplicative renormalization or improvement factors as discussed in Refs.~\cite{Huntley:1986de,Bali:1997am,Koma:2006fw,Guazzini:2007bu,Christensen:2016wdo}.

We define such a multiplicative improvement factor $\ZE$ corresponding to a finite renormalization via 
\begin{eqnarray}
\label{EQN_ZE} \ZE(a) = \frac{F_{\partial V}(r^\ast,a)}{F_E(r^\ast,a)} ,
\end{eqnarray} 
where $r^\ast$ is an arbitrary separation.
After determining this renormalization factor $\ZE(a)$ at a single arbitrary separation $r^\ast$, it can be used to improve $F_E(r,a)$ at all other separations according to
\begin{eqnarray}
F_E^\text{ren}(r,a) = \ZE(a) F_E(r,a).
\end{eqnarray} 
$F_E^\text{ren}(r,a)$ should then have significantly smaller discretization errors than $F_E(r,a)$ 
and, thus, is expected to be quite close to both $F_{\partial V}(r,a)$ and the continuum result $F(r)$.
We note that $\ZE(a) \rightarrow 1$ for  $a \rightarrow 0$.

\begin{figure}[htb]
\begin{center}
\includegraphics[width=0.49\textwidth,page=1]{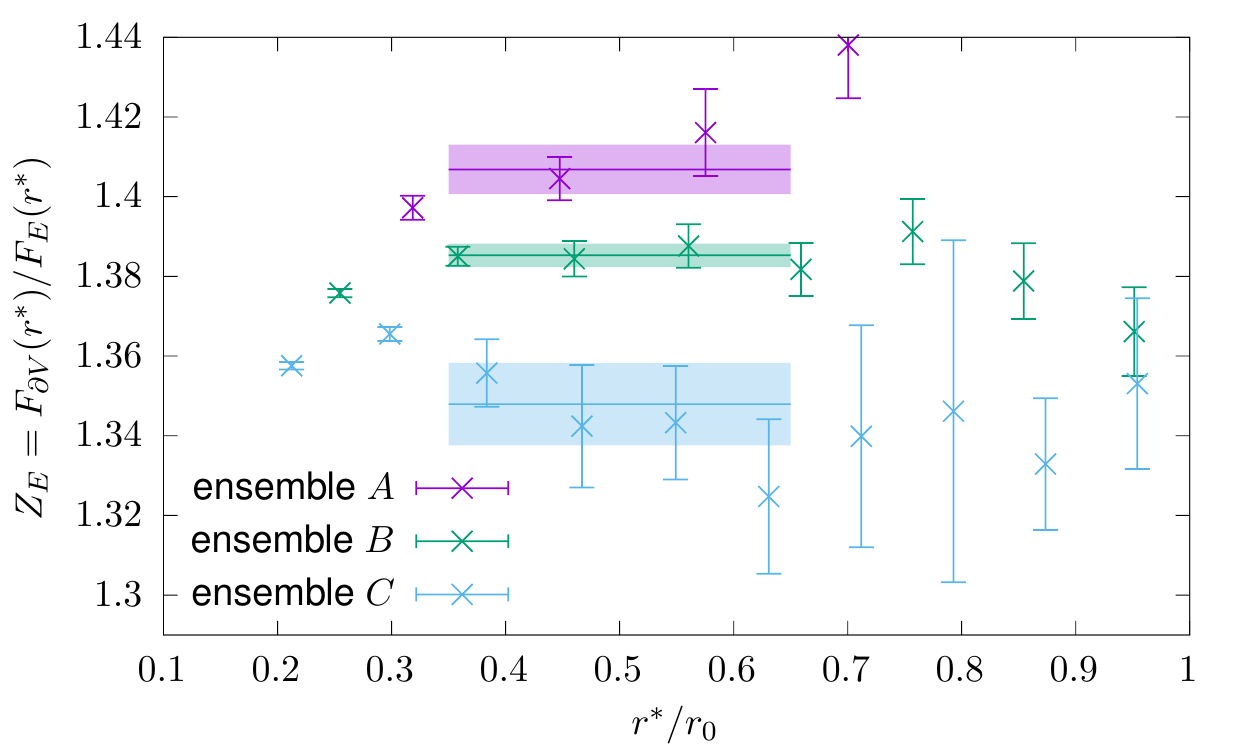}
\includegraphics[width=0.49\textwidth,page=2]{Z_E.pdf}
\end{center}
\caption{\label{FIG002}$\ZE = F_{\partial V} / F_E$ as a function of $r^\ast$. The colored horizontal lines and error bands represent the fits to determine a numerical value for $\ZE$ for each ensemble. \textbf{(left)}~Wilson loops. \textbf{(right)}~Polyakov loops.}
\end{figure}

In Fig.\ \ref{FIG002} we show the renormalization constant $\ZE$, 
defined in Eq.\ \eqref{EQN_ZE}, as a function of $r^\ast$ for both Wilson and Polyakov loops.
The figure exhibits plateau regions that confirm the expected constant behavior of $\ZE$.
Also consistent with expectation is the dependence of $\ZE$ on $\beta$. One can see that with decreasing lattice spacing $a$ the improvement factor $\ZE$ slowly decreases towards $1$.

We determine a numerical value for $\ZE$ for each ensemble by fitting a constant to the lattice data points shown in Fig.\ \ref{FIG002} in the range $0.35 \, r_0 \leq r^\ast \leq 0.65 \, r_0$.
Results of these fits are collected in Table\ \ref{TAB007} for both Wilson and Polyakov loops.
There are small differences between the Wilson loop and the Polyakov loop results, which might be due to different remaining systematic errors.

\begin{table}[htb]
\begin{center}
\def\arraystretch{1.2}
\begin{tabular}{cccc}
\hline
ensemble & $a$ in $\text{fm}$ & $\ZE$ from Wilson loops & $\ZE$ from Polyakov loops \\
\hline
A & $0.060$ & $1.4068(63)$ & $1.4001(20)$ \\
B & $0.048$ & $1.3853(30)$ & $1.3776(10)$ \\
C & $0.040$ & $1.348(11)\phantom{0}$ & $1.3628(13)$ \\
\hline
\end{tabular}
\end{center}
\caption{\label{TAB007}Renormalization constants $\ZE$ obtained by fitting constants to $F_{\partial V} / F_E$ in the range $0.35 \, r_0 \leq r^\ast \leq 0.65 \, r_0$.}
\end{table}

\section{Numerical results for the static force}\label{sec:res}

\begin{figure}[htb]
\begin{center}
\includegraphics[width=0.75\textwidth]{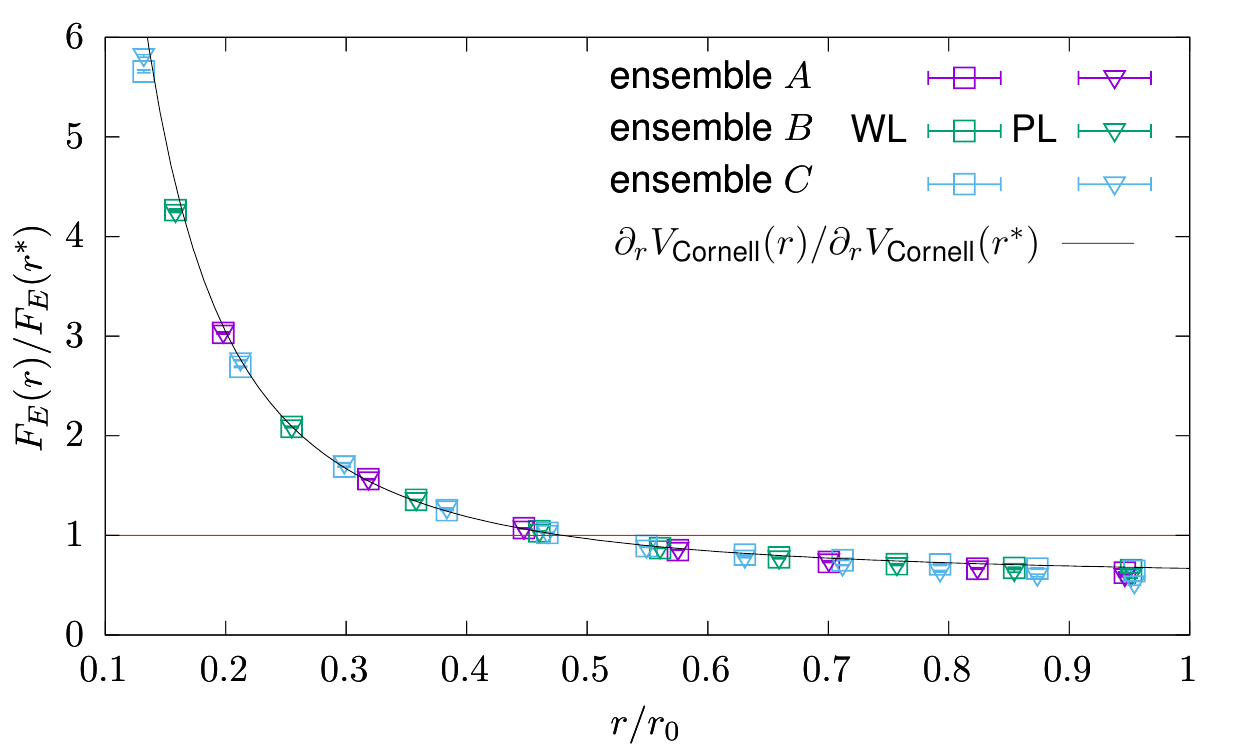}
\end{center}
\caption{\label{FIG003} Left: $F_E(r) / F_E(r^\ast)$ as a function of $r$ for $r^\ast = 0.48 \, r_0 \approx 0.24 \, \textrm{fm}$ obtained from Wilson loops (boxes) and Polyakov loops (triangles). 
For comparison we also show $\partial_r V_\textrm{Cornell}(r) / \partial_r V_\textrm{Cornell}(r^\ast)$.} 
\end{figure}

Now we consider $F_E(r) / F_E(r^\ast)$, where $F_E$ is the non-renormalized static force defined in Eq.\ \eqref{eq:f1}. 
We choose a fixed separation $r^\ast = 0.48 \, r_0 \approx 0.24 \, \textrm{fm}$ such that $r^\ast / a$ is close to an integer for all three ensembles.
Corresponding numerical results for Wilson loops as well as for Polyakov loops are shown in Fig.\ \ref{FIG003}.
For comparison, we also fit a Cornell ansatz $V_\textrm{Cornell}(r) = -\alpha/r + \sigma r$ to the static potential and plot $\partial_r V_\textrm{Cornell}(r) / \partial_r V_\textrm{Cornell}(r^\ast)$.
The agreement of $F_E(r) / F_E(r^\ast)$ and $\partial_r V_\textrm{Cornell}(r) / \partial_r V_\textrm{Cornell}(r^\ast)$ constitutes a numerical proof of concept for the method of computing the static force 
via a chromoelectric field insertion.

In summary, we tested a novel method to compute the static force $F(r)$ from expectation values of Wilson or Polyakov loops with chromoelectric field insertions.
The numerical results exhibit sizable discretization errors and the convergence to the continuum limit is rather slow, 
but this can be compensated by an $r$-independent multiplicative renormalization factor $\ZE$. 
Concerning efficiency, our method appears to be comparable to the traditional method of first computing the static potential and then taking the derivative,
as investigated and discussed in detail in our recent publication \cite{Brambilla:2021wqs}.
We note that the relation between the force
and the color electric field has also been used in recent work from other groups \cite{Baker:2018mhw,Baker:2019gsi} to determine the string tension.

This exploratory computation of the static force is also an important preparatory step for future projects, 
where similar correlation functions
need to be computed. 
An example is the computation of $1/m$ and $1/m^2$ corrections ($m$ denotes the heavy quark mass) to the ordinary static potential or to hybrid static potentials.
We conclude by noting that the renormalization discussed in section~\ref{sec:reno} might not be necessary anymore, when using the gradient flow, 
as discussed during the conference talk. Due to page limitations we do not discuss the gradient flow in the context of the static force in this proceedings contribution, 
but refer to the forthcoming Ref.~\cite{Proc2}.

\FloatBarrier
\acknowledgments{
C.R.~acknowledges support by a Karin and Carlo Giersch Scholarship of the Giersch foundation. M.W.\ acknowledges funding by the Heisenberg Programme of the Deutsche Forschungsgemeinschaft 
(DFG, German Research Foundation) -- Projektnummer 399217702. 
This work has been supported by the NSFC (National Natural Science Foundation of China) and the DFG through the funds provided 
to the Sino-German Collaborative Research Center TRR110 ``Symmetries and the Emergence of Structure in QCD'' (NSFC Grant No.\ 12070131001, DFG Project-ID 196253076 -- TRR 110), and by the DFG cluster of excellence ORIGINS.
Calculations on the GOETHE-HLR and on the FUCHS-CSC high-performance computers of the Frankfurt University were conducted for this research. 
We would like to thank HPC-Hessen, funded by the State Ministry of Higher Education, Research and the Arts, for programming advice. 
Part of the simulations have been carried out on the computing facilities of the Computational Center for Particle and Astrophysics (C2PAP) of the cluster of excellence ORIGINS. 
}

\bibliographystyle{jhep_modified}
\bibliography{proceedings.bib}

\providecommand{\href}[2]{#2}\begingroup\raggedright\begin{thebibliography}{10}

\bibitem{Aoki:2019cca}
{\scshape Flavour Lattice Averaging Group} collaboration, S.~Aoki et~al.,
  \emph{{FLAG Review 2019}},
  \href{https://arxiv.org/abs/1902.08191}{{\ttfamily 1902.08191}}.
\bibitem{Pineda:1998id}
A.~Pineda, \emph{{Heavy quarkonium and nonrelativistic effective field
  theories}}, phd thesis, University of Barcelona, 1998.
\bibitem{Hoang:1998nz}
A.~H. Hoang, M.~C. Smith, T.~Stelzer and S.~Willenbrock, \emph{{Quarkonia and
  the pole mass}},
  \href{https://doi.org/10.1103/PhysRevD.59.114014}{\emph{Phys. Rev. D}
  {\bfseries 59} (1999) 114014}
  [\href{https://arxiv.org/abs/hep-ph/9804227}{{\ttfamily hep-ph/9804227}}].
\bibitem{Necco:2001xg}
S.~Necco and R.~Sommer, \emph{{The $N_f = 0$ heavy quark potential from short
  to intermediate distances}},
  \href{https://doi.org/10.1016/S0550-3213(01)00582-X}{\emph{Nucl. Phys.}
  {\bfseries B622} (2002) 328}
  [\href{https://arxiv.org/abs/hep-lat/0108008}{{\ttfamily hep-lat/0108008}}].
\bibitem{Necco:2001gh}
S.~Necco and R.~Sommer, \emph{{Testing perturbation theory on the $N_f = 0$
  static quark potential}},
  \href{https://doi.org/10.1016/S0370-2693(01)01298-9}{\emph{Phys. Lett. B}
  {\bfseries 523} (2001) 135}
  [\href{https://arxiv.org/abs/hep-ph/0109093}{{\ttfamily hep-ph/0109093}}].
\bibitem{Bazavov:2014soa}
A.~Bazavov, N.~Brambilla, X.~G. Tormo, I, P.~Petreczky, J.~Soto and A.~Vairo,
  \emph{{Determination of $\alpha_s$ from the QCD static energy: An update}},
  \href{https://doi.org/10.1103/PhysRevD.90.074038}{\emph{Phys. Rev. D}
  {\bfseries 90} (2014) 074038}
  [\href{https://arxiv.org/abs/1407.8437}{{\ttfamily 1407.8437}}].
\bibitem{Vairo:2015vgb}
A.~Vairo, \emph{{A low-energy determination of $\alpha_s$ at three loops}},
  \href{https://doi.org/10.1051/epjconf/201612602031,
  10.1051/epjconf/201612007004}{\emph{EPJ Web Conf.} {\bfseries 126} (2016)
  02031} [\href{https://arxiv.org/abs/1512.07571}{{\ttfamily 1512.07571}}].
\bibitem{Vairo:2016pxb}
A.~Vairo, \emph{{Strong coupling from the QCD static energy}},
  \href{https://doi.org/10.1142/S0217732316300391}{\emph{Mod. Phys. Lett.}
  {\bfseries A31} (2016) 1630039}.
\bibitem{Brambilla:2000gk}
N.~Brambilla, A.~Pineda, J.~Soto and A.~Vairo, \emph{{The QCD potential at
  O(1/m)}}, \href{https://doi.org/10.1103/PhysRevD.63.014023}{\emph{Phys. Rev.}
  {\bfseries D63} (2001) 014023}
  [\href{https://arxiv.org/abs/hep-ph/0002250}{{\ttfamily hep-ph/0002250}}].
\bibitem{Brambilla:2021wqs}
N.~Brambilla, V.~Leino, O.~Philipsen, C.~Reisinger, A.~Vairo and M.~Wagner,
  \emph{{Lattice gauge theory computation of the static force}},
  \href{https://arxiv.org/abs/2106.01794}{{\ttfamily 2106.01794}}.
\bibitem{Brambilla:2019zqc}
{\scshape TUMQCD} collaboration, N.~Brambilla, V.~Leino, O.~Philipsen,
  C.~Reisinger, A.~Vairo and M.~Wagner, \emph{{Static force from the lattice}},
  \href{https://doi.org/10.22323/1.363.0109}{\emph{PoS} {\bfseries LATTICE2019}
  (2019) 109} [\href{https://arxiv.org/abs/1911.03290}{{\ttfamily
  1911.03290}}].
\bibitem{Bali:1997am}
G.~S. Bali, K.~Schilling and A.~Wachter, \emph{{Complete $\mathcal{O}(v^2)$
  corrections to the static interquark potential from SU(3) gauge theory}},
  \href{https://doi.org/10.1103/PhysRevD.56.2566}{\emph{Phys. Rev.} {\bfseries
  D56} (1997) 2566} [\href{https://arxiv.org/abs/hep-lat/9703019}{{\ttfamily
  hep-lat/9703019}}].
\bibitem{Luscher:2001up}
M.~Lüscher and P.~Weisz, \emph{{Locality and exponential error reduction in
  numerical lattice gauge theory}},
  \href{https://doi.org/10.1088/1126-6708/2001/09/010}{\emph{JHEP} {\bfseries
  09} (2001) 010} [\href{https://arxiv.org/abs/hep-lat/0108014}{{\ttfamily
  hep-lat/0108014}}].
\bibitem{Jansen:2008si}
{\scshape ETM} collaboration, K.~Jansen, C.~Michael, A.~Shindler and M.~Wagner,
  \emph{{The Static-light meson spectrum from twisted mass lattice QCD}},
  \href{https://doi.org/10.1088/1126-6708/2008/12/058}{\emph{JHEP} {\bfseries
  12} (2008) 058} [\href{https://arxiv.org/abs/0810.1843}{{\ttfamily
  0810.1843}}].
\bibitem{Lepage:1992xa}
G.~P. Lepage and P.~B. Mackenzie, \emph{{On the viability of lattice
  perturbation theory}},
  \href{https://doi.org/10.1103/PhysRevD.48.2250}{\emph{Phys. Rev. D}
  {\bfseries 48} (1993) 2250}
  [\href{https://arxiv.org/abs/hep-lat/9209022}{{\ttfamily hep-lat/9209022}}].
\bibitem{Huntley:1986de}
A.~Huntley and C.~Michael, \emph{{Spin - Spin and Spin - Orbit Potentials From
  Lattice Gauge Theory}},
  \href{https://doi.org/10.1016/0550-3213(87)90438-X}{\emph{Nucl. Phys.}
  {\bfseries B286} (1987) 211}.
\bibitem{Koma:2006fw}
Y.~Koma and M.~Koma, \emph{{Spin-dependent potentials from lattice QCD}},
  \href{https://doi.org/10.1016/j.nuclphysb.2007.01.033}{\emph{Nucl. Phys.}
  {\bfseries B769} (2007) 79}
  [\href{https://arxiv.org/abs/hep-lat/0609078}{{\ttfamily hep-lat/0609078}}].
\bibitem{Guazzini:2007bu}
{\scshape ALPHA} collaboration, D.~Guazzini, H.~B. Meyer and R.~Sommer,
  \emph{{Non-perturbative renormalization of the chromo-magnetic operator in
  Heavy Quark Effective Theory and the B* - B mass splitting}},
  \href{https://doi.org/10.1088/1126-6708/2007/10/081}{\emph{JHEP} {\bfseries
  10} (2007) 081} [\href{https://arxiv.org/abs/0705.1809}{{\ttfamily
  0705.1809}}].
\bibitem{Christensen:2016wdo}
C.~Christensen and M.~Laine, \emph{{Perturbative renormalization of the
  electric field correlator}},
  \href{https://doi.org/10.1016/j.physletb.2016.02.020}{\emph{Phys. Lett. B}
  {\bfseries 755} (2016) 316}
  [\href{https://arxiv.org/abs/1601.01573}{{\ttfamily 1601.01573}}].
\bibitem{Baker:2018mhw}
M.~Baker, P.~Cea, V.~Chelnokov, L.~Cosmai, F.~Cuteri and A.~Papa,
  \emph{{Isolating the confining color field in the SU(3) flux tube}},
  \href{https://doi.org/10.1140/epjc/s10052-019-6978-y}{\emph{Eur. Phys. J. C}
  {\bfseries 79} (2019) 478}
  [\href{https://arxiv.org/abs/1810.07133}{{\ttfamily 1810.07133}}].
\bibitem{Baker:2019gsi}
M.~Baker, P.~Cea, V.~Chelnokov, L.~Cosmai, F.~Cuteri and A.~Papa, \emph{{The
  confining color field in SU(3) gauge theory}},
  \href{https://doi.org/10.1140/epjc/s10052-020-8077-5}{\emph{Eur. Phys. J. C}
  {\bfseries 80} (2020) 514}
  [\href{https://arxiv.org/abs/1912.04739}{{\ttfamily 1912.04739}}].
\bibitem{Proc2}
V.~Leino, N.~Brambilla, J.~Mayer-Steudte and A.~Vairo, \emph{{The static force
  from generalized Wilson loops using gradient flow}}, {\emph{In preparation}
  {\bfseries TUM-EFT 157/21} (2021) }.
\end{thebibliography}\endgroup

\end{document}